\documentclass[runningheads]{llncs}
\usepackage{cite}
\usepackage{graphicx}
\usepackage{subfigure}
\usepackage[font=scriptsize,labelfont=bf]{caption}
\usepackage{floatrow}
\usepackage{pgfplots}
\usepackage{todonotes}
\usepackage{amsmath,amssymb}
\usepackage{array, booktabs, ragged2e}
\usetikzlibrary{positioning,shapes,arrows,patterns,fit,backgrounds,calc,svg.path,3d,decorations.text,shapes.arrows}
\newcommand{\ra}[1]{\renewcommand{\arraystretch}{#1}}

\pgfdeclarelayer{background}
\pgfdeclarelayer{front}
\pgfsetlayers{background,main,front}

\makeatletter
\pgfkeys{%
	/tikz/on layer/.code={
		\pgfonlayer{#1}\begingroup
		\aftergroup\endpgfonlayer
		\aftergroup\endgroup
	},
	/tikz/node on layer/.code={
		\gdef\node@@on@layer{%
			\setbox\tikz@tempbox=\hbox\bgroup\pgfonlayer{#1}\unhbox\tikz@tempbox\endpgfonlayer\egroup}
		\aftergroup\node@on@layer
	},
	/tikz/end node on layer/.code={
		\endpgfonlayer\endgroup\endgroup
	}
}

\def\node@on@layer{\aftergroup\node@@on@layer}

\begin{document}
\title{Quantized Neural Networks for Radar Interference Mitigation}
%
%
\author{Johanna Rock\inst{1} \and
Wolfgang Roth\inst{1} \and
Paul Meissner\inst{2} \and
Franz Pernkopf\inst{1}}
\authorrunning{J. Rock et al.}
%
\institute{Graz University of Technology, Austria \\\email{\{johanna.rock,roth,pernkopf\}@tugraz.at} \and
Infineon Technologies Austria AG, Graz\\
\email{paul.meissner@infineon.com}}
\maketitle              
\begin{abstract}
Radar sensors are crucial for environment perception of driver assistance systems as well as autonomous vehicles. Key performance factors are weather resistance and the possibility to directly measure velocity. With a rising number of radar sensors and the so far unregulated automotive radar frequency band, mutual interference is inevitable and must be dealt with. Algorithms and models operating on radar data in early processing stages are required to run directly on specialized hardware, i.e. the radar sensor. This specialized hardware typically has strict resource-constraints, i.e. a low memory capacity and low computational power.
Convolutional Neural Network (CNN)-based approaches for denoising and interference mitigation yield promising results for radar processing in terms of performance. However, these models typically contain millions of parameters, stored in hundreds of megabytes of memory, and require additional memory during execution.
In this paper we investigate quantization techniques for CNN-based denoising and interference mitigation of radar signals. We analyze the quantization potential of different CNN-based model architectures and sizes by considering (i) quantized weights and (ii) piecewise constant activation functions, which results in reduced memory requirements for model storage and during the inference step respectively.

\keywords{Quantization Aware Training \and Resource-Efficiency \and Binarized Convolutional Neural Networks \and Straight Through Estimator \and Interference Mitigation \and Automotive Radar}
\end{abstract}
\section{Introduction}

\emph{Advanced Driver Assistance Systems (ADAS)} and \emph{Autonomous Vehicles (AV)} heavily rely on a multitude of heterogeneous  sensors for environment perception. Among them are radar sensors, that are used for object detection, classification and to directly measure relative object velocities. Advantages of radar sensors are a high resolution, their robustness concerning difficult weather and lighting conditions, as well as their capability to directly measure the relative object velocity.

Typically \emph{frequency modulated continuous wave (FMCW)/chirp sequence (CS)} radars are used in the automotive context. They transmit sequences of linearly modulated radio frequency (RF) chirps on a shared and non-regulated band. This may lead to mutual interference of multiple radar sensors, becoming increasingly likely with higher numbers of deployed radar-enhanced vehicles and higher bandwidths due to better range-resolutions.

For a non-regulated spectrum, the most common form of mutual interference is non-coherent, where the transmitters send with non-identical parameters ~\cite{TOT18}. This results in burst-like interferences in time domain and a decreased detection sensitivity in the range-Doppler (RD) map. Thus, the detection and mitigation of interference is crucial in a safety context and must be addressed.

Several conventional signal processing algorithms for interference mitigation of mutual interference have been proposed. The most simplistic method is to substitute all interference-affected samples with zero, followed by an optional smoothing of the boundaries. More advanced methods use nonlinear filtering in slow-time~\cite{WAG18}, iterative reconstruction using Fourier transforms and thresholding~\cite{MAR12}, estimation and subtraction of the interference component~\cite{BEC17}, or beamforming~\cite{Bechter2016}.
Recently, the use of deep learning has been proposed for radar spectra denoising and interference mitigation. For this task neural networks are applied in time domain or in frequency domain, typically in a supervised manner. For interference mitigation in time domain, recurrent neural networks (RNNs) are used in \cite{8690848, 9053013}. Basic CNN-based models in \cite{9114627} and U-Net inspired CNNs in \cite{9114641} are applied to frequency domain signals. While the results are impressive on simulated and measurement data, the problem of high memory and computational requirements by these models has not been addressed in detail. In order to use the aforementioned methods for interference mitigation in practice, they have to comply with memory, computational as well as real-time constraints of specialized hardware, i.e. the radar sensor.

Typically, deep neural networks (DNNs) have thousands or even millions of parameters and require hundreds of megabytes memory to be stored and executed. Note, that memory is often the limiting factor also in terms of energy efficiency and execution time, because loading data dominates over arithmetic operations and loading from off-chip DRAM is magnitudes more costly than accessing data from on-chip SRAM \cite{han2015learning}.

There are several, partly orthogonal, options to reduce memory and computational requirements \cite{roth2020resourceefficient}. The initial network architecture contributes substantially to the resource requirements, thus a small model with few parameters and small feature-maps is preferable. \emph{Neural architecture search (NAS)} can be applied with resource-oriented objectives in order to find efficient models \cite{cai2018proxylessnas}. Other approaches are different network pruning techniques, weight sharing, knowledge distillation, special matrix structures and quantization. In a quantized neural network, weights and activations, hence feature-maps, are discretized and thus their bit-width is reduced. Typically, research on neural network quantization considers standard image classification data sets (e.g. MNIST, CIFAR-10 or ImageNet) rather than real-world data or regression tasks.

In this paper we investigate the suitability of quantization techniques, in particular the \emph{Straight Through Estimator (STE)} \cite{NIPS20166573}, to reduce the total memory requirements for interference mitigation on radar sensors using CNN-based models from \cite{9114627}. In our experiments we use real-world FMCW/CS radar measurements with simulated interference. Main contributions of this paper are:
\begin{itemize}
	\item We analyze binarization capabilities according to different model architectures, sizes and quantization strategies, i.e. quantized weights, activations or both
	\item We illustrate the importance of resource-efficient real-valued initial models w.r.t. quantization and the resulting memory requirements
	\item We present preliminary results for quantizing exceptionally small models without significant performance degradation
\end{itemize}

\section{Signal model}

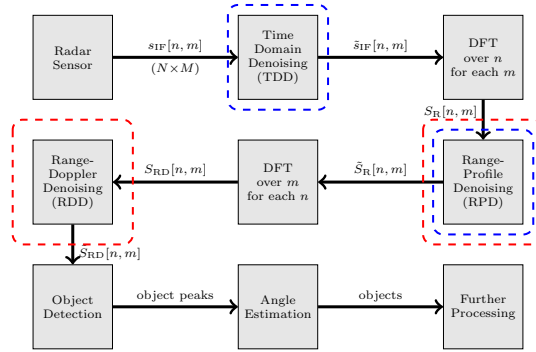
\begin{figure}
	\centering
	\scriptsize
	\resizebox{7.8cm}{!}{
		\tikzstyle{block}=[draw, fill=black!10, text width=5em, text centered, minimum height=6em]
\tikzstyle{blockfocus}=[block, draw=red, thick, dashed,rounded corners]
\tikzstyle{blockinactive}=[block, opacity=.3]

\tikzstyle{arrow}=[very thick, ->]
\tikzstyle{arrowinactive}=[arrow, opacity=.2]

\begin{tikzpicture}[scale=0.7, transform shape]
\begin{scope}[node distance=0.8cm and 2.5cm]

\node[block] (dft2) at (0,0) {DFT over $m$ \\ for each $n$};
\node[block, above left = of dft2] (rs) {Radar Sensor};
\node[block, above = of dft2] (tdp) {Time Domain Denoising (TDD)};
\node[block, above right = of dft2] (dft1) {DFT over $n$ \\ for each $m$};
\node[block, left = of dft2] (rdd) {Range-Doppler Denoising (RDD)};
\node[block, right = of dft2] (rpd) {Range-Profile Denoising (RPD)};
\node[block, below left = of dft2] (od) {Object Detection};
\node[block, below = of dft2] (ae) {Angle Estimation};
\node[block, below right = of dft2] (fp) {Further Processing};

\draw[arrow] (rs) -- (tdp) node[midway, above] () {$s_{\mathrm{IF}}[n,m]$} node [midway,below] () {${(N{\times}M)}$};
\draw[arrow] (tdp) -- (dft1) node[midway, above] () {$\tilde{s}_{\mathrm{IF}}[n,m]$};
\draw[arrow] (dft1) -- (rpd) node[pos=0.3, left] () {$S_{\mathrm{R}}[n,m]$};
\draw[arrow] (rpd) -- (dft2) node[midway, above] () {$\tilde{S}_{\mathrm{R}}[n,m]$};
\draw[arrow] (dft2) -- (rdd) node[midway, above] () {$S_{\mathrm{RD}}[n,m]$};
\draw[arrow] (rdd) -- (od) node[pos=0.7, right] () {$\tilde{S}_{\mathrm{RD}}[n,m]$};
\draw[arrow] (od) -- (ae) node[midway, above] () {object peaks};
\draw[arrow] (ae) -- (fp) node[midway, above] () {objects};

\draw[red,thick,dashed,rounded corners] ($(rpd.north west)+(-0.4,0.4)$)  rectangle ($(rpd.south east)+(0.4,-0.4)$);
\draw[red,thick,dashed,rounded corners] ($(rdd.north west)+(-0.4,0.4)$)  rectangle ($(rdd.south east)+(0.4,-0.4)$);

\draw[blue,thick,dashed,rounded corners] ($(tdp.north west)+(-0.2,0.2)$)  rectangle ($(tdp.south east)+(0.2,-0.2)$);
\draw[blue,thick,dashed,rounded corners] ($(rpd.north west)+(-0.2,0.2)$)  rectangle ($(rpd.south east)+(0.2,-0.2)$);

\end{scope}
\end{tikzpicture}
	}
	\caption{Block diagram of a basic FMCW/CS radar processing chain. Dashed boxes indicate the locations of optional interference mitigation steps, including CNN-based approaches (red) and classical methods (blue).}
	\label{fig:spchain_classical}
\end{figure}

The \emph{range-Doppler (RD) processing} chain of a common FMCW/CS radar is depicted in Fig.~\ref{fig:spchain_classical}. The radar sensor transmits a set of linearly modulated RF chirps, also termed ramps. Object reflections are perceived by the receive antennas and mixed with the transmit signal resulting in the \emph{Intermediate Frequency (IF) Signal}. The objects' distances and velocities are contained in the sinusoidals' frequencies and their linear phase change over successive ramps~\cite{STO92,WIN07}, respectively.
The signal is processed as a $N \times M$ data matrix $s_{\mathrm{IF}}[n,m]$, containing $N$ fast time samples within one ramp and $M$ ramps. Discrete Fourier transforms (DFTs) are computed over both dimensions, yielding a two-dimensional spectrum, the RD map $S_{\mathrm{RD}}[n,m]$, on which peaks can be found at positions corresponding to the objects' distances and velocities. After peak detection, further processing can include angular estimation, tracking, and classification.

State-of-the-art (``classical") interference mitigation methods are mostly signal processing algorithms that are applied either on the time domain signal $s_{\mathrm{IF}}[n,m]$ or on the frequency domain signal $S_{\mathrm{R}}[n,m]$ after the first DFT \cite{TOT18}. The CNN-based method used in this paper, also denoted \emph{Range-Doppler Denoising (RDD)\label{rd-denoising}}, is applied on the RD map after the second DFT.

\section{CNN Model}
\begin{figure*}
	\centering
	\scriptsize
	\includegraphics[width=0.9\textwidth]{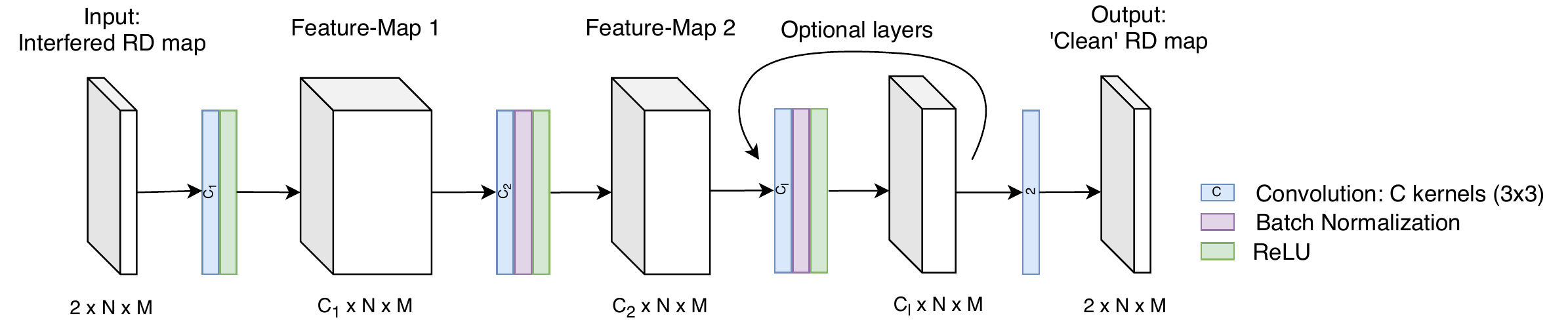}
	\caption{CNN architecture for radar signal denoising and interference mitigation. It uses ReLu, \emph{Batch Normalization (BN)} and the convolution operation $\textrm{Conv}(i, o, (s_1\times s_2))$, for $i$ input channels, $o$ output channels, and a kernel size of $s_{1} \times s_{2}$. See Figures~\ref{fig:cnn_archA} and ~\ref{fig:cnn_archB} for the two concrete model variants used in this paper.}
	\label{fig:cnn_arch}
	\vspace{-2mm}
\end{figure*}

The model architecture is based on \cite{9114627} and is illustrated in Figure \ref{fig:cnn_arch}. The network contains $L$ layers, each being a composite function of operations including the convolution operation (Conv), ReLu activation function \cite{journals/jmlr/GlorotBB11} and \emph{Batch Normalization (BN)}. The last layer uses a linear activation function and two feature-maps corresponding to the real and imaginary values of the complex-valued network output. From a signal processing perspective, the CNN model filters the RD map using learnable filter kernels.

\begin{figure}
	\centering
	\scriptsize
	\linespread{0.6}
	\subfigure[\scriptsize Architecture A: same number of channels in all layers]{
		\centering
		\includegraphics[width=0.45\textwidth]{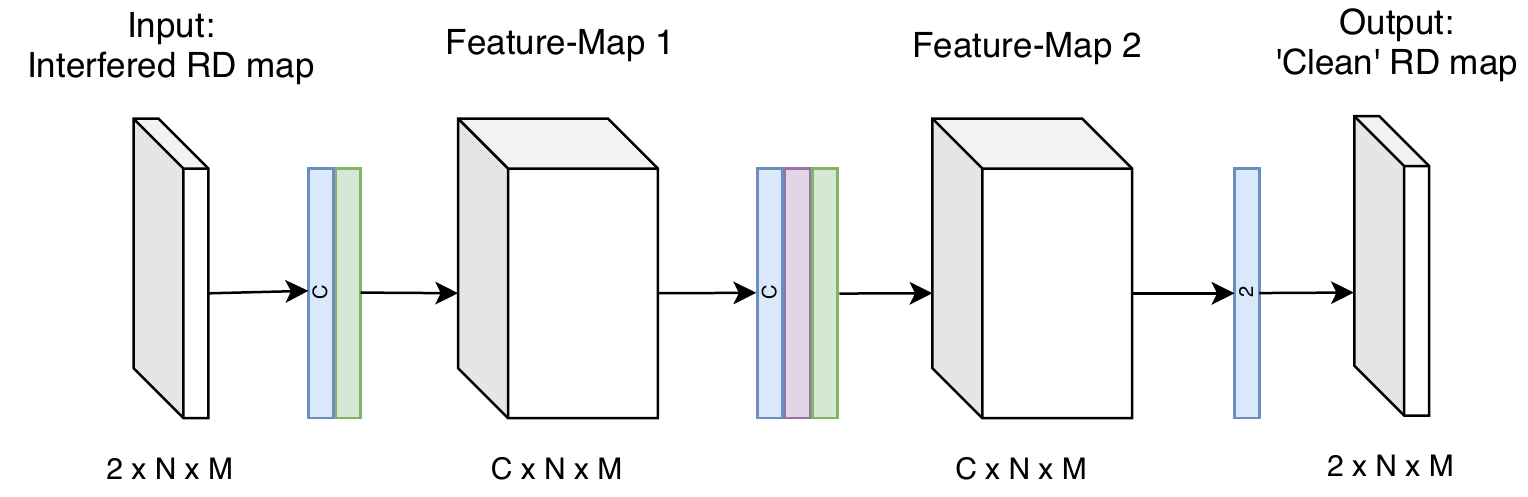}
		\label{fig:cnn_archA}
	}
\hspace{5mm}
	\subfigure[\scriptsize Architecture B: bottle-neck based architecture of channels, i.e. the number of channels C is halved for each layer]{
		\centering
		\includegraphics[width=0.45\textwidth]{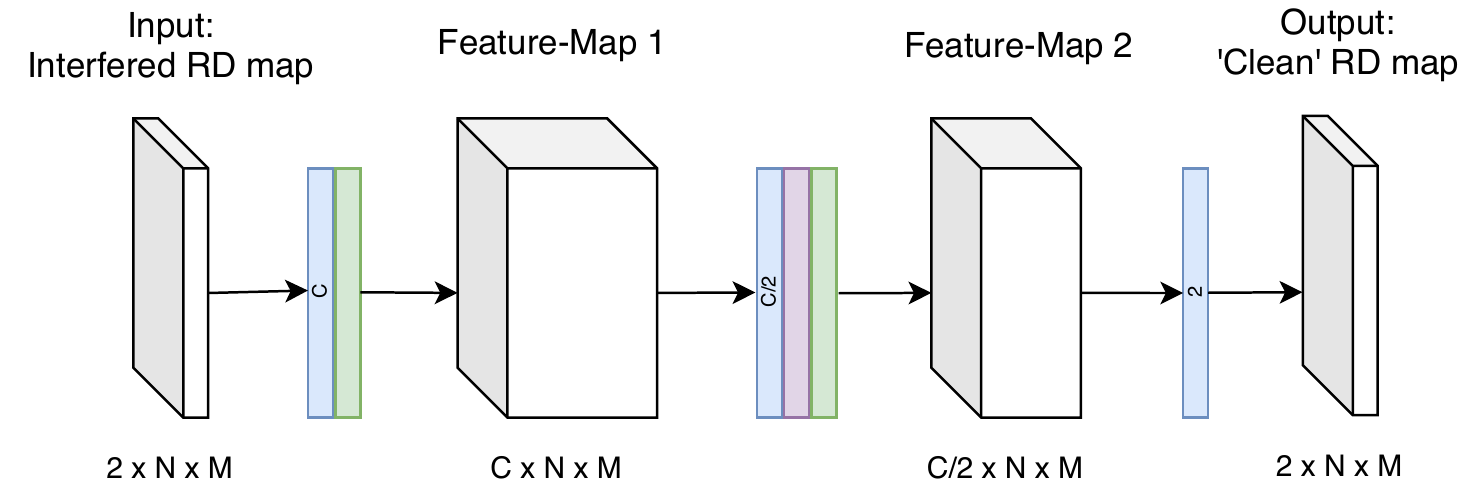}
		\label{fig:cnn_archB}
	}
	\caption{Two concrete variants of the CNN-based model architecture as used in this paper. Both architectures are depicted for an exemplary model with $L=3$ layers and $C$ channels.}
\end{figure}

The model is applied to radar snapshots after the second DFT (RD maps), hence the input samples are complex valued patches of size $N \times M$. We use two input channels in order to represent the real- and imaginary parts of the complex valued input. The network inputs are RD maps with interference and their targets are the corresponding 'clean' RD maps without interference. Square kernels are used in combination with zero-padding, such that the inputs and outputs for each layer have the same spatial dimension. For the training of the network we use the \emph{mean squared error (MSE)} loss function and  the \emph{Adam} algorithm \cite{DBLP:journals/corr/KingmaB14}.
In this paper, we report results for two different variants of the CNN-based model:

\begin{itemize}
	\item[A] \label{sec:modelarchA} consists of the same number of channels in every layer, except for the output layer which has a fixed number of two channels. A model denoted as L3\_C32\_A would thus have three layers with C = [32, 32, 2] channels. See Figure~\ref{fig:cnn_archA}.
	\item[B] \label{sec:modelarchB} has a bottleneck-based structure of channels where the number of channels is halved for each layer and the last layer always consists of two channels. A model denoted as L3\_C32\_B would thus have three layers with C = [32, 16, 2] channels. See Figure~\ref{fig:cnn_archB}.
\end{itemize}

\section{Quantization}

The training of real-valued neural networks is typically performed using gradient-based algorithms that update the network weights according to some loss function. Discrete-valued weights or piecewise constant activation functions incorporate non-differentiable components, whose gradient is zero almost everywhere, such that conventional gradient-based optimization is not possible. Quantization can be achieved by:
\begin{enumerate}
	\item Quantizing pre-trained real-valued neural networks in a more or less heuristic manner, e.g. rounding of weights.
	\item Quantization aware training using real-valued auxiliary weights and the \emph{Straight Through Gradient Estimator (STE)} during the backward pass of quantization functions \cite{NIPS20166573}.
	\item Training \emph{weight distributions over discrete weights} using a Bayesian inspired approach. The most probable weights of the trained network can be chosen in order to obtain the discrete-valued NN \cite{5d1d891bcd074f37974c1f34cafb6e8e}.
\end{enumerate}
In this paper we consider trained quantization of weights and activations using the STE.

\subsection{Straight Through Gradient Estimator (STE)}

\begin{figure}
	\centering
	\footnotesize
	\floatbox[{\capbeside\thisfloatsetup{capbesideposition={right,center},capbesidewidth=0.48\textwidth}}]{figure}[\FBwidth]
	{\caption{Computation of forward (red) and backward (green) pass through a simplified NN building block using the straight through gradient estimator (STE). The building block consists of a convolution with quantized weights $W_q^l$ followed by a sign activation function. Q denotes the piecewise constant quantization function; in the forward pass it is applied to the real-valued auxiliary weights $W^l$. During backpropagation the green dashed line is followed, where the zero-gradients are avoided and substituted by the gradients of the tanh and identity respectively. The gradient updates are then applied to the real-valued weights $W^l$ according to the gradient based learning algorithm.}\label{fig:ste}}
	{\includegraphics[width=0.48\textwidth]{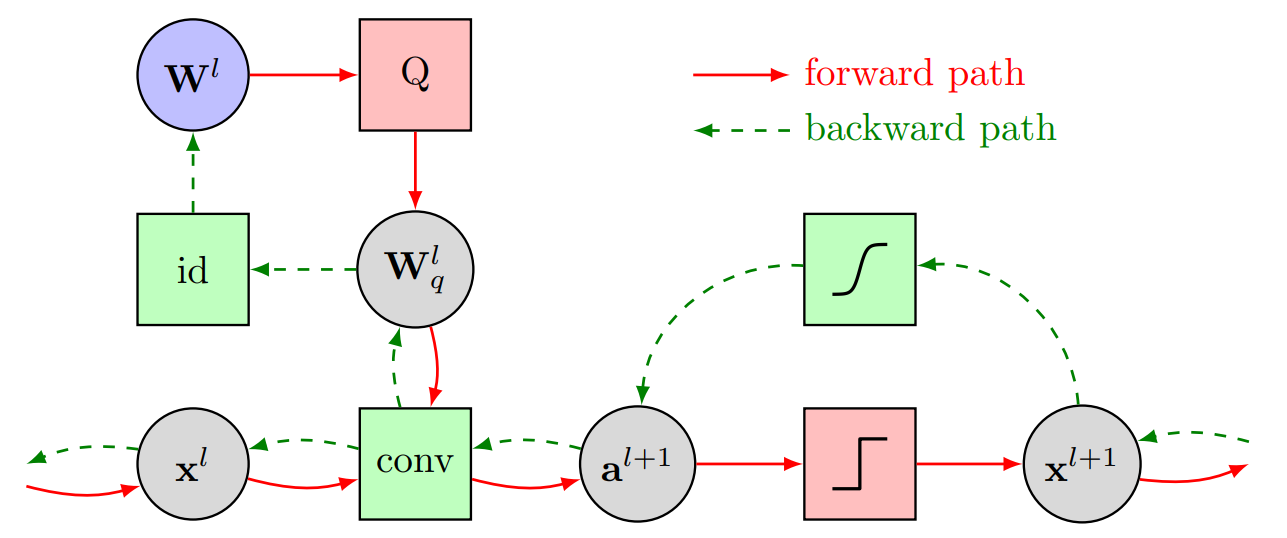}}
\end{figure}

The STE is a simple method for gradient approximation of non-differentiable network components, that achieves remarkable results in practice. Essentially, the gradient of zero-gradient quantization functions is replaced by some non-zero value during the backward pass of the network. Let $f(w)$ be some non-differentiable function within the computation graph of the loss function $\mathcal{L}$ such that the partial derivative $\frac{\partial \mathcal{L}}{\partial w}$ is not defined. The gradient $\frac{\partial \mathcal{L}}{\partial w}$ is then approximated by the STE using

\begin{equation}
\frac{\partial \mathcal{L}}{\partial w} = \frac{\partial \mathcal{L}}{\partial f} \frac{\partial f}{\partial w} \approx \frac{\partial \mathcal{L}}{\partial f} \frac{\partial \tilde{f}}{\partial w},
\end{equation}
where $\tilde{f}(w)$ is an arbitrary differentiable function. Typically, $\tilde{f}(w)$ is chosen to have a similar functional shape as $f(w)$ or to be the identity $\tilde{f}(w)=w$, which results in a derivative of $\tilde{f}'(w)=1$ and thus passes the gradient on to higher components in the computation graph. Figure~\ref{fig:ste} illustrates the computation graph of the STE on a simplified NN layer consisting of a convolution using quantized weights and the sign activation function.

In this paper, we consider two different quantization functions:
\begin{itemize}
	\item \textbf{Binary:} sign
	\[
	Q_B(x)= 
	\begin{cases}
	+1,	& \text{if } x\geq 0\\
	-1,	& \text{if } x < 0
	\end{cases}
	\]
	\item \textbf{Integer quantization:} rounding
	\[
	Q_I(x)= \textrm{round}(x),
	\]
	where $\textrm{round}(x)$ denotes the rounding of $x$ to the closest integer value that is representable with a specified number of bits.
\end{itemize}
The \emph{dynamic range} can be used to map discrete weights to a real-valued range of values with a simple multiplication. Hence, the discrete model weights are stored alongside one real-valued number per layer, i.e. the dynamic range. These discrete weights are scaled according to that value, which typically boosts the model performance. Note, that the memory requirements for the dynamic range can be neglected, because only one 32 bit value is stored per layer. The dynamic range can be seen as a scaling factor $\alpha$, such that
\[
W_q = Q(W / \alpha) \alpha,
\]
where $W_q$ are the quantized weights, $W$ are the real-valued auxiliary weights and $Q$ is the quantization function.

\section{Experimental Setup}
In this paper, we use real FMCW/CS radar measurement data combined with simulated interference to obtain input-output pairs for training CNN models in order to perform the denoising and interference mitigation tasks. The model is applied to the processed radar signal after the second DFT, i.e. the RD map. The overall goal is the correct detection of peaks in the RD map, that correspond to real objects rather than clutter or noise.

\subsection{Data set}
The data set used in this paper consists of real-world FMCW/CS radar measurements mixed with simulated interference. The measurements were recorded in typical inner-city scenarios, where each measurement consists of 32 consecutive radar snapshots (RD maps) captured with sixteen antennas. The radar signal consists of reflections from static and moving objects as well as receiver noise. The interference is simulated by sampling uniformly from the ego radar, interferer radar and signal processing parameters, and is added to the time domain measurement signal. See \cite{Rock1907:Complex} for a listing and detailed description of the simulation parameters and~\cite{9114627, toth2020analysis} for an extensive analysis of the used measurement signals.

\subsection{Evaluation}
The \emph{F1-Score} is used as evaluation metric, it is defined as:

\begin{equation}
	\mathrm{F_1} = 2 \frac{
		\mathrm{precision}  \cdot \mathrm{recall}
	} {
		\mathrm{precision} + \mathrm{recall}
	}
\end{equation}
We use manual labels that were obtained from the clean measurement RD maps without interference as ground truth target detections. A \emph{Cell Averaging Constant False Alarm Rate (CA-CFAR)} target detection algorithm \cite{scharf1991statistical} is used to automatically extract detections, hence peak locations, from the interference mitigated model outputs. The ground truth target detections and the CA-CFAR generated detections from interference mitigated RD maps are the basis for the F1-Score. The F1-Score comprises the harmonic mean of correct detections and false alarms. All evaluation results are reported as the mean and the standard deviation over three individually trained models if not stated otherwise.

\section{Experimental Results}

First, we analyze the overall suitability of CNN-based models from \cite{9114627} to be discretized without noteworthy performance degradation.

\begin{figure}
	\centering
	\scriptsize
	\subfigure[\scriptsize Layers]{
		\resizebox {0.5\textwidth} {!} {
\begin{tikzpicture}

\definecolor{color0}{rgb}{0.172549019607843,0.627450980392157,0.172549019607843}
\definecolor{color1}{rgb}{0.83921568627451,0.152941176470588,0.156862745098039}
\definecolor{color2}{rgb}{0.580392156862745,0.403921568627451,0.741176470588235}
\definecolor{color3}{rgb}{0.12156862745098,0.466666666666667,0.705882352941177}
\definecolor{color4}{rgb}{1,0.498039215686275,0.0549019607843137}

\begin{axis}[
scale only axis=true,
width=0.5\columnwidth,
height=0.35\columnwidth,
axis line style={white},
tick align=outside,
tick pos=left,
xlabel={Model},
xmajorgrids,
xmin=0.9, xmax=3.1,
xtick style={color=white!33.3333333333333!black},
xtick={1,2,3},
xticklabels={L3-C1024,L5-C1024,L7-C1024},
ylabel={F1-Score},
ymajorgrids,
ymin=0.829394605429512, ymax=0.912609964357914,
ytick style={color=white!33.3333333333333!black},
ytick={0.82,0.83,0.84,0.85,0.86,0.87,0.88,0.89,0.9,0.91,0.92},
yticklabels={0.82,0.83,0.84,0.85,0.86,0.87,0.88,0.89,0.90,0.91,0.92}
]
\path [draw=color0, fill=color0, opacity=0.2, very thin]
(axis cs:1,0.902897931604162)
--(axis cs:1,0.902897931604162)
--(axis cs:2,0.897604626763606)
--(axis cs:3,0.895901738534557)
--(axis cs:3,0.895901738534557)
--(axis cs:3,0.895901738534557)
--(axis cs:2,0.899118919805831)
--(axis cs:1,0.902897931604162)
--cycle;

\path [draw=color0, fill=color0, opacity=0.2, very thin]
(axis cs:1,0.902012058200272)
--(axis cs:1,0.900161424338658)
--(axis cs:2,0.896518099271671)
--(axis cs:3,0.896609350358013)
--(axis cs:3,0.898222277521526)
--(axis cs:3,0.898222277521526)
--(axis cs:2,0.899814448360316)
--(axis cs:1,0.902012058200272)
--cycle;

\path [draw=color1, fill=color1, opacity=0.2, very thin]
(axis cs:1,0.878822097602364)
--(axis cs:1,0.83317712174444)
--(axis cs:2,0.875582841001602)
--(axis cs:3,0.871494150197411)
--(axis cs:3,0.895419171185435)
--(axis cs:3,0.895419171185435)
--(axis cs:2,0.901396377151305)
--(axis cs:1,0.878822097602364)
--cycle;

\path [draw=color1, fill=color1, opacity=0.2, very thin]
(axis cs:1,0.879031305846938)
--(axis cs:1,0.852180134208572)
--(axis cs:2,0.885677187310018)
--(axis cs:3,0.883533569189472)
--(axis cs:3,0.899205849234899)
--(axis cs:3,0.899205849234899)
--(axis cs:2,0.89541217428892)
--(axis cs:1,0.879031305846938)
--cycle;

\path [draw=color2, fill=color2, opacity=0.2, very thin]
(axis cs:1,0.852695209947125)
--(axis cs:1,0.849829478455929)
--(axis cs:2,0.868192703504482)
--(axis cs:3,0.878408221125369)
--(axis cs:3,0.881791624885775)
--(axis cs:3,0.881791624885775)
--(axis cs:2,0.87301867410171)
--(axis cs:1,0.852695209947125)
--cycle;

\path [draw=color2, fill=color2, opacity=0.2, very thin]
(axis cs:1,0.84775907187048)
--(axis cs:1,0.846150111420446)
--(axis cs:2,0.853640943481085)
--(axis cs:3,0.8440879047742)
--(axis cs:3,0.846243028984415)
--(axis cs:3,0.846243028984415)
--(axis cs:2,0.856995651994883)
--(axis cs:1,0.84775907187048)
--cycle;

\addplot [semithick, color3]
table {%
0.9 0.908827448042986
3.1 0.908827448042986
};
\addplot [semithick, color4]
table {%
0.9 0.836633657524875
3.1 0.836633657524875
};
\addplot [semithick, color0, dashed]
table {%
1 0.902897931604162
2 0.898361773284719
3 0.895901738534557
};
\addplot [semithick, color0]
table {%
1 0.901086741269465
2 0.898166273815993
3 0.89741581393977
};
\addplot [semithick, color1, dashed]
table {%
1 0.855999609673402
2 0.888489609076454
3 0.883456660691423
};
\addplot [semithick, color1]
table {%
1 0.865605720027755
2 0.890544680799469
3 0.891369709212185
};
\addplot [semithick, color2, dashed]
table {%
1 0.851262344201527
2 0.870605688803096
3 0.880099923005572
};
\addplot [semithick, color2]
table {%
1 0.846954591645463
2 0.855318297737984
3 0.845165466879307
};
\end{axis}

\end{tikzpicture}
		}
		\label{fig:result1a}
	}
	\hspace{-0.9cm}
	\subfigure[\scriptsize Channels]{
		\resizebox {0.5\textwidth} {!} {
\begin{tikzpicture}

\definecolor{color0}{rgb}{0.172549019607843,0.627450980392157,0.172549019607843}
\definecolor{color1}{rgb}{0.83921568627451,0.152941176470588,0.156862745098039}
\definecolor{color2}{rgb}{0.580392156862745,0.403921568627451,0.741176470588235}
\definecolor{color3}{rgb}{0.12156862745098,0.466666666666667,0.705882352941177}
\definecolor{color4}{rgb}{1,0.498039215686275,0.0549019607843137}

\begin{axis}[
scale only axis=true,
width=0.5\columnwidth,
height=0.35\columnwidth,
axis line style={white},
legend cell align={left},
legend style={fill opacity=0.8, draw opacity=1, text opacity=1, at={(0.97,0.03)}, anchor=south east, draw=white!80!black, font=\scriptsize},
tick align=outside,
tick pos=left,
xlabel={Model},
xmajorgrids,
xmin=0.9, xmax=3.1,
xtick style={color=white!33.3333333333333!black},
xtick={1,2,3},
xticklabels={L7-C32,L7-C256,L7-C1024},
ylabel={F1-Score},
ymajorgrids,
ymin=0.1749988745532, ymax=0.943771665828214,
ytick style={color=white!33.3333333333333!black},
ytick={0.1,0.2,0.3,0.4,0.5,0.6,0.7,0.8,0.9,1},
yticklabels={0.1,0.2,0.3,0.4,0.5,0.6,0.7,0.8,0.9,1.0}
]
\path [draw=color0, fill=color0, opacity=0.2, very thin]
(axis cs:1,0.898125065444409)
--(axis cs:1,0.897022575320038)
--(axis cs:2,0.896074539518329)
--(axis cs:3,0.895901738534557)
--(axis cs:3,0.895901738534557)
--(axis cs:3,0.895901738534557)
--(axis cs:2,0.899093199399536)
--(axis cs:1,0.898125065444409)
--cycle;

\path [draw=color0, fill=color0, opacity=0.2, very thin]
(axis cs:1,0.893321047732215)
--(axis cs:1,0.887081011773133)
--(axis cs:2,0.896502161533129)
--(axis cs:3,0.896609350358013)
--(axis cs:3,0.898222277521526)
--(axis cs:3,0.898222277521526)
--(axis cs:2,0.898363015516298)
--(axis cs:1,0.893321047732215)
--cycle;

\path [draw=color1, fill=color1, opacity=0.2, very thin]
(axis cs:1,0.884531466968341)
--(axis cs:1,0.857011232870117)
--(axis cs:2,0.874306752140633)
--(axis cs:3,0.871494150197411)
--(axis cs:3,0.895419171185435)
--(axis cs:3,0.895419171185435)
--(axis cs:2,0.890142676510463)
--(axis cs:1,0.884531466968341)
--cycle;

\path [draw=color1, fill=color1, opacity=0.2, very thin]
(axis cs:1,0.879528282110347)
--(axis cs:1,0.233961731863795)
--(axis cs:2,0.847323625628789)
--(axis cs:3,0.883533569189472)
--(axis cs:3,0.899205849234899)
--(axis cs:3,0.899205849234899)
--(axis cs:2,0.885687626967746)
--(axis cs:1,0.879528282110347)
--cycle;

\path [draw=color2, fill=color2, opacity=0.2, very thin]
(axis cs:1,0.802513271207556)
--(axis cs:1,0.797376403467372)
--(axis cs:2,0.857168406827616)
--(axis cs:3,0.878408221125369)
--(axis cs:3,0.881791624885775)
--(axis cs:3,0.881791624885775)
--(axis cs:2,0.862502987873036)
--(axis cs:1,0.802513271207556)
--cycle;

\path [draw=color2, fill=color2, opacity=0.2, very thin]
(axis cs:1,0.436100883897742)
--(axis cs:1,0.209943092338428)
--(axis cs:2,0.79558615278744)
--(axis cs:3,0.8440879047742)
--(axis cs:3,0.846243028984415)
--(axis cs:3,0.846243028984415)
--(axis cs:2,0.80447677362606)
--(axis cs:1,0.436100883897742)
--cycle;

\addplot [semithick, color0, dashed]
table {%
1 0.897573820382224
2 0.897583869458933
3 0.895901738534557
};
\addlegendentry{Real A}
\addplot [semithick, color0]
table {%
1 0.890201029752674
2 0.897432588524714
3 0.89741581393977
};
\addlegendentry{Real B}
\addplot [semithick, color1, dashed]
table {%
1 0.870771349919229
2 0.882224714325548
3 0.883456660691423
};
\addlegendentry{Binary A}
\addplot [semithick, color1]
table {%
1 0.556745006987071
2 0.866505626298268
3 0.891369709212185
};
\addlegendentry{Binary B}
\addplot [semithick, color2, dashed]
table {%
1 0.799944837337464
2 0.859835697350326
3 0.880099923005572
};
\addlegendentry{Sign A}
\addplot [semithick, color2]
table {%
1 0.323021988118085
2 0.80003146320675
3 0.845165466879307
};
\addlegendentry{Sign B}
\addplot [semithick, color3]
table {%
	0.9 0.908827448042987
	3.1 0.908827448042987
};
\addlegendentry{Clean}
\addplot [semithick, color4]
table {%
	0.9 0.836633657524875
	3.1 0.836633657524875
};
\addlegendentry{Interfered}
\end{axis}

\end{tikzpicture}
		}
		\label{fig:result1b}
	}
	\caption{Performance comparison of binarized models using different numbers of layers \ref{fig:result1a} and channels \ref{fig:result1b}. For each model configuration on the x-axis, the solid line indicates the performance of a bottleneck-based architecture of channels (Architecture B, see Section~\ref{sec:modelarchB}) and the dashed line indicates the performance of an architecture with the same number of channels, i.e. 1024, in each layer (Architecture A, see Section~\ref{sec:modelarchB}).}
	\label{fig:result1ab}
	\vspace{-2mm}
\end{figure}
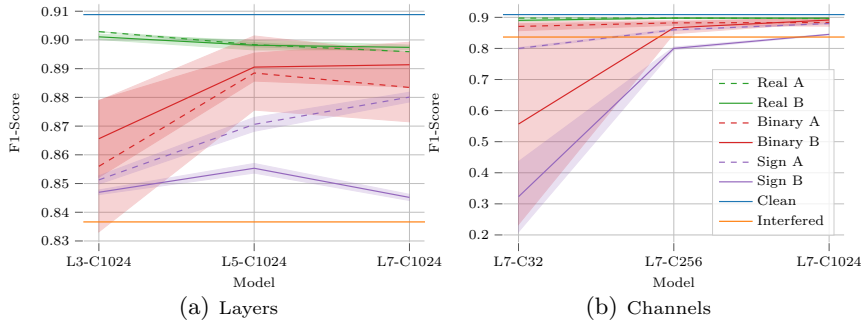
Figure \ref{fig:result1ab} shows a comparison of model architectures A and B with the same number of maximal channels (i.e. 1024 channels), and different numbers of layers (i.e. 3, 5, and 7 layers). The F1-Score of the clean measurement data (Clean) and the interfered data (Interfered) are indicated as references; note, that only a model surpassing the score for interfered data yields an improvement.

The real-valued baseline (Real) in Figure \ref{fig:result1a} does not strongly depend on the model architecture (A or B) within certain limits of layers and channels. Models with binary weights (Binary) or activations (Sign) typically yield better results with a higher number of model parameters. Architecture B yields better results for binarized weights whereas architecture A is better suited for binarized activations. For architecture B the limiting factor is the minimal number of channels, i.e. the number of channels in the $L-1^{th}$ layer for a model with $L$ layers. Hence, the overall number of parameters as well as the minimal number of channels has a strong impact on the model performance.

Figure \ref{fig:result1b} shows a comparison between models with the same number of layers (i.e. 7 layers) and different numbers of channels (i.e. 32, 256 and 1024 channels). Binary weight (Binary) and binary activation (Sign) models depend highly on the number of channels. Not only the minimal number of channels is a limiting factor, but also the total number of channels, which is shown by comparing architecture B with 32 and 256 channels, where they both have 8 channels in layer $L-1$ but the model with an overall higher number of channels performs better. Models with binary activations require a very large number of channels, and thus parameters, in order to reach a high F1-Score.

In summary, we have shown that binary weight models can almost reach the performance of their real-valued equivalent given a high number of model parameters and especially channels. In the binary case, architecture B is preferable. For binary activations however, architecture A performs better. In any case, a large amount of parameters is required in order to reach a high F1-Score.

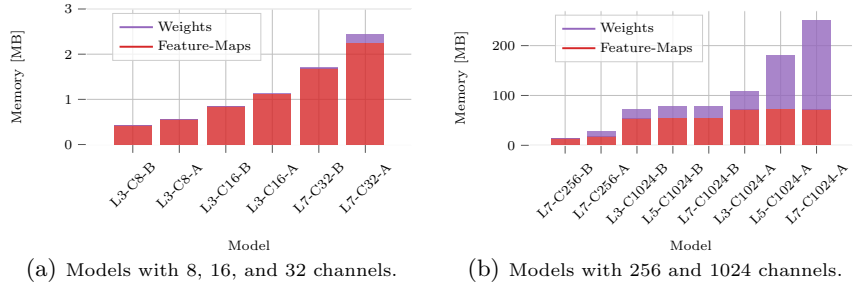
\begin{figure*}
	\centering
	\scriptsize
	\subfigure[\scriptsize Models with 8, 16, and 32 channels.]{
		\resizebox {0.46\textwidth} {!} {
\begin{tikzpicture}

\definecolor{color0}{rgb}{0.83921568627451,0.152941176470588,0.156862745098039}  
\definecolor{color1}{rgb}{0.580392156862745,0.403921568627451,0.741176470588235}  

\begin{axis}[
scale only axis=true,
width=0.48\textwidth,
height=0.2\textwidth,
axis line style={white},
legend cell align={left},
legend style={fill opacity=0.8, draw opacity=1, text opacity=1, draw=white!80!black, at={(0.08,0.97)}, anchor=north west},
legend entries={{Weights},{Feature-Maps}},
tick align=outside,
tick pos=left,
xlabel={Model},
x label style={at={(axis description cs:.5,-0.45)}},
xmajorgrids,
xmin=-1, xmax=6,
xtick style={color=white!33.3333333333333!black},
xtick={0,1,2,3,4,5},
xticklabel style = {rotate=45.0},
xticklabels={L3-C8-B,L3-C8-A,L3-C16-B,L3-C16-A,L7-C32-B,L7-C32-A},
ylabel={Memory [MB]},
ymajorgrids,
ymin=0, ymax=3,
ytick style={color=white!33.3333333333333!black}
]

\addlegendimage{solid, no markers, color1, line width=1pt}
\addlegendimage{solid, no markers, color0, line width=1pt}

\draw[draw=none,fill=color0,fill opacity=0.8,very thin] (axis cs:-0.4,0) rectangle (axis cs:0.4,0.421875);
\draw[draw=none,fill=color0,fill opacity=0.8,very thin] (axis cs:0.6,0) rectangle (axis cs:1.4,0.5625);
\draw[draw=none,fill=color0,fill opacity=0.8,very thin] (axis cs:1.6,0) rectangle (axis cs:2.4,0.84375);
\draw[draw=none,fill=color0,fill opacity=0.8,very thin] (axis cs:2.6,0) rectangle (axis cs:3.4,1.125);
\draw[draw=none,fill=color0,fill opacity=0.8,very thin] (axis cs:3.6,0) rectangle (axis cs:4.4,1.6875);
\draw[draw=none,fill=color0,fill opacity=0.8,very thin] (axis cs:4.6,0) rectangle (axis cs:5.4,2.25);
\draw[draw=none,fill=color1,fill opacity=0.8,very thin] (axis cs:-0.4,0.421875) rectangle (axis cs:0.4,0.423797607421875);
\draw[draw=none,fill=color1,fill opacity=0.8,very thin] (axis cs:0.6,0.5625) rectangle (axis cs:1.4,0.5657958984375);
\draw[draw=none,fill=color1,fill opacity=0.8,very thin] (axis cs:1.6,0.84375) rectangle (axis cs:2.4,0.84979248046875);
\draw[draw=none,fill=color1,fill opacity=0.8,very thin] (axis cs:2.6,1.125) rectangle (axis cs:3.4,1.135986328125);
\draw[draw=none,fill=color1,fill opacity=0.8,very thin] (axis cs:3.6,1.6875) rectangle (axis cs:4.4,1.71881103515625);
\draw[draw=none,fill=color1,fill opacity=0.8,very thin] (axis cs:4.6,2.25) rectangle (axis cs:5.4,2.43017578125);
\end{axis}
\end{tikzpicture}
		}
		\label{fig:result_mem1}
	}
	\subfigure[\scriptsize Models with 256 and 1024 channels.]{
		\resizebox {0.46\textwidth} {!} {
\begin{tikzpicture}

\definecolor{color0}{rgb}{0.83921568627451,0.152941176470588,0.156862745098039}  
\definecolor{color1}{rgb}{0.580392156862745,0.403921568627451,0.741176470588235}  

\begin{axis}[
scale only axis=true,
width=0.48\textwidth,
height=0.2\textwidth,
axis line style={white},
legend cell align={left},
legend style={fill opacity=0.8, draw opacity=1, text opacity=1, draw=white!80!black, at={(0.08,0.97)}, anchor=north west},
legend entries={{Weights},{Feature-Maps}},
tick align=outside,
tick pos=left,
xlabel={Model},
x label style={at={(axis description cs:.5,-0.45)}},
xmajorgrids,
xmin=5, xmax=14,
xtick style={color=white!33.3333333333333!black},
xtick={6,7,8,9,10,11,12,13},
xticklabel style = {rotate=45.0},
xticklabels={L7-C256-B,L7-C256-A,L3-C1024-B,L5-C1024-B,L7-C1024-B,L3-C1024-A,L5-C1024-A,L7-C1024-A},
ylabel={Memory [MB]},
ymajorgrids,
ymin=0, ymax=270,
ytick style={color=white!33.3333333333333!black}
]

\addlegendimage{solid, no markers, color1, line width=1pt}
\addlegendimage{solid, no markers, color0, line width=1pt}

\draw[draw=none,fill=color0,fill opacity=0.8,very thin] (axis cs:5.6,0) rectangle (axis cs:6.4,13.5);
\draw[draw=none,fill=color0,fill opacity=0.8,very thin] (axis cs:6.6,0) rectangle (axis cs:7.4,18);
\draw[draw=none,fill=color0,fill opacity=0.8,very thin] (axis cs:7.6,0) rectangle (axis cs:8.4,54);
\draw[draw=none,fill=color0,fill opacity=0.8,very thin] (axis cs:8.6,0) rectangle (axis cs:9.4,54);
\draw[draw=none,fill=color0,fill opacity=0.8,very thin] (axis cs:9.6,0) rectangle (axis cs:10.4,54);
\draw[draw=none,fill=color0,fill opacity=0.8,very thin] (axis cs:10.6,0) rectangle (axis cs:11.4,72);
\draw[draw=none,fill=color0,fill opacity=0.8,very thin] (axis cs:11.6,0) rectangle (axis cs:12.4,72);
\draw[draw=none,fill=color0,fill opacity=0.8,very thin] (axis cs:12.6,0) rectangle (axis cs:13.4,72);
\draw[draw=none,fill=color1,fill opacity=0.8,very thin] (axis cs:5.6,13.5) rectangle (axis cs:6.4,15.0166625976562);
\draw[draw=none,fill=color1,fill opacity=0.8,very thin] (axis cs:6.6,18) rectangle (axis cs:7.4,29.28515625);
\draw[draw=none,fill=color1,fill opacity=0.8,very thin] (axis cs:7.6,54) rectangle (axis cs:8.4,72.10546875);
\draw[draw=none,fill=color1,fill opacity=0.8,very thin] (axis cs:8.6,54) rectangle (axis cs:9.4,77.7041015625);
\draw[draw=none,fill=color1,fill opacity=0.8,very thin] (axis cs:9.6,54) rectangle (axis cs:10.4,78.049072265625);
\draw[draw=none,fill=color1,fill opacity=0.8,very thin] (axis cs:10.6,72) rectangle (axis cs:11.4,108.140625);
\draw[draw=none,fill=color1,fill opacity=0.8,very thin] (axis cs:11.6,72) rectangle (axis cs:12.4,180.140625);
\draw[draw=none,fill=color1,fill opacity=0.8,very thin] (axis cs:12.6,72) rectangle (axis cs:13.4,252.140625);
\end{axis}
\end{tikzpicture}
		}
		\label{fig:result_mem2}
	}
	\caption{Total memory requirements for real-valued models during the inference step stated in megabytes. Purple indicates memory requirements for weights and red for the two largest consecutive feature-maps. The model names indicate their architecture using the scheme L\textless LAYERS\textgreater -C\textless CHANNELS\textgreater -\{A/B\}, where A or B indicates architecture A or B.}
	\label{fig:result_mem}
	\vspace{-2mm}
\end{figure*}

Memory requirements during the inference step stem from storing (i) model parameters and (ii) temporary results during the computation, i.e. feature-maps. For the sake of run time and energy efficient computing, the model parameters and two consecutive feature-maps have to be stored in fast accessible on-chip memory simultaneously. Hence, the memory requirement is given by the memory to store the model parameters and to store the two consecutive feature-maps with the highest accumulated memory requirements.

Figure~\ref{fig:result_mem} shows the total memory requirements per model architecture. All depicted models are real-valued and reach a similar F1-Score of $F1 \ge 0.89$. Models with few channels (e.g. 8, 16 or 32) have \emph{much smaller} memory requirements than models with many channels (e.g. 256 or 1024); note the different y-axis scales in Figures~\ref{fig:result_mem1} and \ref{fig:result_mem2}. Quantization however reduces the memory footprint by a factor of up to 32, i.e. in the binary case. Thus, there is only a small subset of real-valued models depicted in Figure~\ref{fig:result_mem} that could be used as base models in order to further reduce memory requirements using quantization.

\begin{figure*}
	{
		\centering
		\scriptsize
		\resizebox{\textwidth}{!}{
\begin{tikzpicture}

\definecolor{color0}{rgb}{0.172549019607843,0.627450980392157,0.172549019607843}
\definecolor{color1}{rgb}{0.83921568627451,0.152941176470588,0.156862745098039}
\definecolor{color2}{rgb}{0.580392156862745,0.403921568627451,0.741176470588235}
\definecolor{color3}{rgb}{0.12156862745098,0.466666666666667,0.705882352941177}
\definecolor{color4}{rgb}{1,0.498039215686275,0.0549019607843137}

\begin{axis}[
scale only axis=true,
width=0.9\textwidth,
height=0.35\textwidth,
legend cell align={left},
legend style={fill opacity=0.8, draw opacity=1, text opacity=1, draw=white!80!black, 
	at={(0.02,0.41)}, anchor=north west},
legend entries={{Real},{Binary},{Sign},{Clean},{Interfered}},
axis line style={white},
log basis x={10},
tick align=outside,
tick pos=left,
xlabel={Memory [MB]},
xmajorgrids,
xmin=0.423881530761719, xmax=200,
xmode=log,
xtick style={color=white!33.3333333333333!black},
xtick={0.01,0.1,1,10,100,1000,10000},
xticklabels={\(\displaystyle {10^{-2}}\),\(\displaystyle {10^{-1}}\),\(\displaystyle {10^{0}}\),\(\displaystyle {10^{1}}\),\(\displaystyle {10^{2}}\),\(\displaystyle {10^{3}}\),\(\displaystyle {10^{4}}\)},
ylabel={F1-Score},
ymajorgrids,
ymin=0.825, ymax=0.925,
ytick style={color=white!33.3333333333333!black},
ytick={0.82,0.84,0.86,0.88,0.9,0.92,0.94},
yticklabels={0.82,0.84,0.86,0.88,0.90,0.92,0.94}
]

\addlegendimage{only marks, mark=*, draw=color0,fill=color0}
\addlegendimage{only marks, mark=*, draw=color1,fill=color1}
\addlegendimage{only marks, mark=*, draw=color2,fill=color2}
\addlegendimage{draw=color3,semithick}
\addlegendimage{draw=color4,semithick}

\addplot [scatter, only marks, mark=*, scatter/use mapped color={
	draw=color0,
	fill=color0,
}, fill=color0, fill opacity=0.7, colormap/viridis, visualization depends on={\thisrow{sizedata} \as\perpointmarksize}, scatter/@pre marker code/.append style={/tikz/mark size=\perpointmarksize}]
table{%
x                      y                      sizedata
1.71949005126953 0.890201029752674 3.46593742488784
2.43213653564453 0.897573820382224 5.34482333212864
15.0204849243164 0.897432588524714 9.08476217547867
29.3007888793945 0.897583869458933 14.9959309468778
78.064338684082 0.89741581393977 18.1157411545607
252.203132629395 0.895901738534557 29.9655737576612
72.1152420043945 0.901086741269465 16.8746688314898
108.156257629395 0.902897931604162 20.0561088512119
77.7182693481445 0.898166273815993 18.050277724276
180.179695129395 0.898361773284719 26.3767688320872
0.849952697753906 0.89602647761059 2.30815716391312
1.13623809814453 0.899994953380168 2.67119521399816
0.423881530761719 0.888376159413645 1.74799081460642
0.565925598144531 0.895253786576051 1.9900830119242
};
\addplot [scatter, only marks, mark=*, scatter/use mapped color={
	draw=color0,
	fill=color0,
},  fill=color0, fill opacity=1, visualization depends on={\thisrow{sizedata} \as\perpointmarksize}, scatter/@pre marker code/.append style={/tikz/mark size=\perpointmarksize}]
table{%
	x                      y                      sizedata
	1.71949005126953 0.890201029752674 0.4
	2.43213653564453 0.897573820382224 0.4
	15.0204849243164 0.897432588524714 0.4
	29.3007888793945 0.897583869458933 0.4
	78.064338684082 0.89741581393977 0.4
	252.203132629395 0.895901738534557 0.4
	72.1152420043945 0.901086741269465 0.4
	108.156257629395 0.902897931604162 0.4
	77.7182693481445 0.898166273815993 0.4
	180.179695129395 0.898361773284719 0.4
	0.849952697753906 0.89602647761059 0.4
	1.13623809814453 0.899994953380168 0.4
	0.423881530761719 0.888376159413645 0.4
	0.565925598144531 0.895253786576051 0.4
};
\addplot [scatter, only marks, mark=*, scatter/use mapped color={
	draw=color1,
	fill=color1,
}, fill=color1, fill opacity=0.7, colormap/viridis, visualization depends on={\thisrow{sizedata} \as\perpointmarksize}, scatter/@pre marker code/.append style={/tikz/mark size=\perpointmarksize}]
table{%
x                      y                      sizedata
1.68915748596191 0.556745006987071 3.46593742488784
2.25759124755859 0.870771349919229 5.34482333212864
13.5512180328369 0.866505626298268 9.08476217547867
18.368293762207 0.882224714325548 14.9959309468778
54.7667999267578 0.891369709212185 18.1157411545607
77.6919021606445 0.883456660691423 29.9655737576612
54.6096267700195 0.865605720027755 16.8746688314898
73.2131423950195 0.855999609673402 20.0561088512119
54.7549209594727 0.890544680799469 18.050277724276
0.844631195068359 0.772597789906985 2.30815716391312
};
\addplot [scatter, only marks, mark=*, scatter/use mapped color={
	draw=color1,
	fill=color1,
}, fill=color1, fill opacity=1, visualization depends on={\thisrow{sizedata} \as\perpointmarksize}, scatter/@pre marker code/.append style={/tikz/mark size=\perpointmarksize}]
table{%
	x y sizedata
	1.68915748596191 0.556745006987071 0.4
	2.25759124755859 0.870771349919229 0.4
	13.5512180328369 0.866505626298268 0.4
	18.368293762207 0.882224714325548 0.4
	54.7667999267578 0.891369709212185 0.4
	77.6919021606445 0.883456660691423 0.4
	54.6096267700195 0.865605720027755 0.4
	73.2131423950195 0.855999609673402 0.4
	54.7549209594727 0.890544680799469 0.4
	0.844631195068359 0.772597789906985 0.4
};
\addplot [scatter, only marks, mark=*, scatter/use mapped color={
	draw=color2,
	fill=color2,
}, fill=color2, fill opacity=0.7, colormap/viridis, visualization depends on={\thisrow{sizedata} \as\perpointmarksize}, scatter/@pre marker code/.append style={/tikz/mark size=\perpointmarksize}]
table{%
x                      y                      sizedata
0.137458801269531 0.323021988118085 3.46593742488784
0.287605285644531 0.799944837337464 5.34482333212864
1.94235992431641 0.80003146320675 9.08476217547867
11.8632888793945 0.859835697350326 14.9959309468778
25.751838684082 0.845165466879307 18.1157411545607
182.453132629395 0.880099923005572 29.9655737576612
19.8027420043945 0.846954591645463 16.8746688314898
38.4062576293945 0.851262344201527 20.0561088512119
25.4057693481445 0.855318297737984 18.050277724276
110.429695129395 0.870605688803096 26.3767688320872
};
\addplot [scatter, only marks, mark=*, scatter/use mapped color={
	draw=color2,
	fill=color2,
}, fill=color2, fill opacity=1, colormap/viridis,visualization depends on={\thisrow{sizedata} \as \perpointmarksize}, scatter/@pre marker code/.append style={/tikz/mark size=\perpointmarksize}]
table {
	x                      y                      sizedata
	0.137458801269531 0.323021988118085 0.4
	0.287605285644531 0.799944837337464 0.4
	1.94235992431641 0.80003146320675 0.4
	11.8632888793945 0.859835697350326 0.4
	25.751838684082 0.845165466879307 0.4
	182.453132629395 0.880099923005572 0.4
	19.8027420043945 0.846954591645463 0.4
	38.4062576293945 0.851262344201527 0.4
	25.4057693481445 0.855318297737984 0.4
	110.429695129395 0.870605688803096 0.4
};
\addplot [semithick, color3]
table {%
0.423881530761719 0.908827448042987
200 0.908827448042987
};
\addplot [semithick, color4]
table {%
0.423881530761719 0.836633657524875
200 0.836633657524875
};
\addplot [scatter, thick, only marks, mark=o, scatter/use mapped color={
	draw=black,
},  color=black, colormap/viridis, visualization depends on={\thisrow{sizedata} \as \perpointmarksize}, scatter/@pre marker code/.append style={/tikz/mark size=\perpointmarksize}]
table{%
	x                      y                      sizedata
	0.423881530761719 0.888376159413645 1.74799081460642
	0.565925598144531 0.895253786576051 1.9900830119242
	0.849952697753906 0.89602647761059 2.30815716391312
	1.13623809814453 0.899994953380168 2.67119521399816
	72.1152420043945 0.901086741269465 16.8746688314898
	108.156257629395 0.902897931604162 20.0561088512119
};
\draw (axis cs:0.423881530761719,0.884) node[
  scale=0.6,
  anchor=base west,
  text=black,
  rotate=0.0,
  node on layer=front,
]{L3-C8-B};
\draw (axis cs:0.5,0.898) node[
  scale=0.6,
  anchor=base west,
  text=black,
  rotate=0.0,
  node on layer=front,
]{L3-C8-A};
\draw (axis cs:0.849952697753906,0.89) node[
  scale=0.6,
  anchor=base west,
  text=black,
  rotate=0.0,
  node on layer=front,
]{L3-C16-B};
\draw (axis cs:1.13623809814453,0.902) node[
  scale=0.6,
  anchor=base west,
  text=black,
  rotate=0.0,
  node on layer=front,
]{L3-C16-A};
\draw (axis cs:2.5,0.870771349919229) node[
  scale=0.6,
  anchor=base west,
  text=black,
  rotate=0.0,
  node on layer=front,
]{L7-C32-A};
\draw (axis cs:13.5512180328369,0.866505626298268) node[
  scale=0.6,
  anchor=base west,
  text=black,
  rotate=0.0,
  node on layer=front,
]{L7-C256-B};
\draw (axis cs:11.8,0.855) node[
  scale=0.6,
  anchor=base west,
  text=black,
  rotate=0.0,
  node on layer=front,
]{L7-C256-A};
\end{axis}

\end{tikzpicture}
		}
		\caption{Average F1-Score vs. memory requirements in megabytes during the inference step of real-valued (Real), binary weight (Binary) and binary activation (Sign) models. The circle volume corresponds to the number of operations required during the inference step. The Pareto optimal points are marked using black borderlines; they all belong to real-valued models. The smallest models for each category are annotated using the format L\textless LAYERS\textgreater -C\textless CHANNELS\textgreater -\{A/B\}, where A or B indicates architecture A or B. See Table \ref{tab:1c} for details of annotated models.}
		\label{fig:result1c}
		\vspace{-2mm}
	}
\end{figure*}
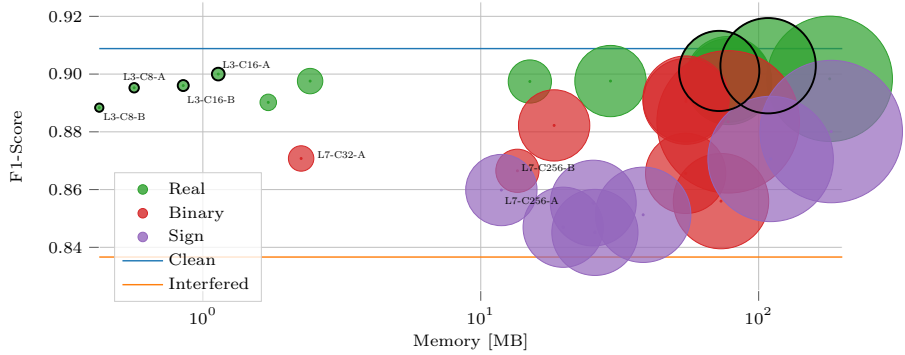

\begin{table*}
	\begin{center}
		\scriptsize
		\caption{Memory and performance details for models annotated in Figure~\ref{fig:result1c}. Model designation consists of L layers, C channels, and architecture A or B. Quantization type Q can be R for real-valued models, B for binary weights, or S for binary activations (sign).}
		\label{tab:1c}
		\ra{1.05}
			\begin{tabular}{lcrrrcccrcrcrcrcr}\toprule
				\multicolumn{5}{c}{Model} & \phantom{abc} & \multicolumn{1}{c}{Q} & \phantom{abc}& \multicolumn{1}{c}{Weights} &
				\phantom{abc} & \multicolumn{1}{c}{Feature-Maps} & \phantom{abc} & \multicolumn{1}{c}{Total} & \phantom{abc} & \multicolumn{1}{c}{Operations} & \phantom{abc} & \multicolumn{1}{c}{F1}\\
				\cmidrule{1-5}
				Name && L & C & A &&&& [MB] && [MB] && [MB] &&$ * 10^{6}$ &&\\
				\midrule
				L3-C8-B && 3 & 8 & B && R && $0.002$ && $0.42$ && $0.42$ && $5$ && $0.8884$ \\
				L3-C8-A && 3 & 8 & A && R && $0.003$ && $0.56$ && $0.57$ && $8$ && $0.8953$ \\
				L3-C16-B && 3 & 16 & B && R && $0.006$ && $0.84$ && $0.85$ && $15$ && $0.8960$ \\
				L3-C16-A && 3 & 16 & A && R && $0.011$ && $1.12$ && $1.14$ && $27$ && $0.9000$ \\
				L7-C32-A && 7 & 32 & A && B && $0.006$ && $2.25$ && $2.26$ && $440$ && $0.8708$ \\
				L7-C256-B && 7 & 256 & B && B && $0.047$ && $13.50$ && $13.55$ && $3678$ && $0.8665$ \\
				L7-C256-A && 7 & 256 & A && S && $11.285$ && $0.56$ && $11.86$ && $27306$ && $0.8598$ \\
				\bottomrule
			\end{tabular}
			\vspace{-5mm}
	\end{center}
\end{table*}

Figure~\ref{fig:result1c} illustrates the performance to memory relation for different quantization types, i.e. real-valued, binary weights, or binary activations. Table \ref{tab:1c} lists details of the smallest models per quantization type. The results clearly show that models with binarized weights or activations reaching an acceptable F1-Score require more memory than a real-valued alternative with fewer parameters. All Pareto optimal points correspond to real-valued models. Already particularly small real-valued models, e.g. the model with three layers and C=[16,8,2] channels, reach a high F1-Score of $F1 > 0.89$.

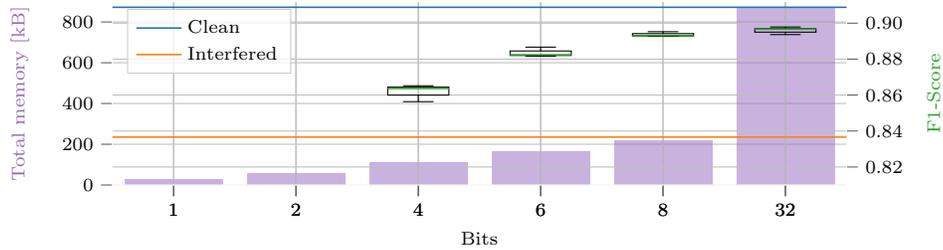
\begin{figure*}
	{
		\centering
		\scriptsize
\begin{tikzpicture}

\definecolor{color0}{rgb}{0.580392156862745,0.403921568627451,0.741176470588235}
\definecolor{color1}{rgb}{0.172549019607843,0.627450980392157,0.172549019607843}
\definecolor{color2}{rgb}{0.12156862745098,0.466666666666667,0.705882352941177}
\definecolor{color3}{rgb}{1,0.498039215686275,0.0549019607843137}

\begin{axis}[
scale only axis=true,
width=0.8\textwidth,
height=0.2\textwidth,
axis line style={white},
tick align=outside,
tick pos=left,
xlabel={Bits},
xmajorgrids,
xmin=0.5, xmax=6.5,
xtick style={color=white!33.3333333333333!black,font=\tiny},
xtick={1,2,3,4,5,6},
xticklabels={1,2,4,6,8,32},
ylabel={Total memory [kB]},
ylabel style={color=color0},
ymajorgrids,
ymin=0, ymax=900,
ytick style={color=white!33.3333333333333!black}
]
\draw[draw=none,fill=color0,fill opacity=0.5,very thin] (axis cs:0.6,0) rectangle (axis cs:1.4,27.357421875);
\draw[draw=none,fill=color0,fill opacity=0.5,very thin] (axis cs:1.6,0) rectangle (axis cs:2.4,54.55078125);
\draw[draw=none,fill=color0,fill opacity=0.5,very thin] (axis cs:2.6,0) rectangle (axis cs:3.4,108.9375);
\draw[draw=none,fill=color0,fill opacity=0.5,very thin] (axis cs:3.6,0) rectangle (axis cs:4.4,163.32421875);
\draw[draw=none,fill=color0,fill opacity=0.5,very thin] (axis cs:4.6,0) rectangle (axis cs:5.4,217.7109375);
\draw[draw=none,fill=color0,fill opacity=0.5,very thin] (axis cs:5.6,0) rectangle (axis cs:6.4,870.3515625);
\end{axis}

\begin{axis}[
scale only axis=true,
width=0.8\textwidth,
height=0.2\textwidth,
axis line style={white},
axis y line=right,
legend cell align={left},
legend style={fill opacity=0.8, draw opacity=1, text opacity=1, at={(0.02,0.96)}, anchor=north west, draw=white!80!black},
tick align=outside,
xmajorgrids,
xmin=0.5, xmax=6.5,
xtick pos=left,
xtick style={color=white!33.3333333333333!black},
xtick={1,2,3,4,5,6},
xticklabels={1,2,4,6,8,32},
ylabel={F1-Score},
ylabel style={color=color1},
ymajorgrids,
ymin=0.81, ymax=0.912,
ytick pos=right,
ytick style={color=white!33.3333333333333!black},
ytick={0.8,0.82,0.84,0.86,0.88,0.9,0.92},
yticklabels={0.80,0.82,0.84,0.86,0.88,0.90,0.92}
]
\addplot [black, forget plot]
table {%
1 0
1 0
};
\addplot [black, forget plot]
table {%
1 0
1 0
};
\addplot [black, forget plot]
table {%
0.875 0
1.125 0
};
\addplot [black, forget plot]
table {%
0.875 0
1.125 0
};
\addplot [black, forget plot]
table {%
2 0.724102604931597
2 0.724067482158539
};
\addplot [black, forget plot]
table {%
2 0.727548151004027
2 0.7309585743034
};
\addplot [black, forget plot]
table {%
1.875 0.724067482158539
2.125 0.724067482158539
};
\addplot [black, forget plot]
table {%
1.875 0.7309585743034
2.125 0.7309585743034
};
\addplot [black, forget plot]
table {%
3 0.860000051320365
3 0.856341720640426
};
\addplot [black, forget plot]
table {%
3 0.864417985222416
3 0.865177588444528
};
\addplot [black, forget plot]
table {%
2.875 0.856341720640426
3.125 0.856341720640426
};
\addplot [black, forget plot]
table {%
2.875 0.865177588444528
3.125 0.865177588444528
};
\addplot [black, forget plot]
table {%
4 0.882029383878705
4 0.881664347471184
};
\addplot [black, forget plot]
table {%
4 0.88452925871811
4 0.886664097149992
};
\addplot [black, forget plot]
table {%
3.875 0.881664347471184
4.125 0.881664347471184
};
\addplot [black, forget plot]
table {%
3.875 0.886664097149992
4.125 0.886664097149992
};
\addplot [black, forget plot]
table {%
5 0.892895973099222
5 0.89266161782335
};
\addplot [black, forget plot]
table {%
5 0.894190336083145
5 0.895250343791197
};
\addplot [black, forget plot]
table {%
4.875 0.89266161782335
5.125 0.89266161782335
};
\addplot [black, forget plot]
table {%
4.875 0.895250343791197
5.125 0.895250343791197
};
\addplot [black, forget plot]
table {%
6 0.894994219860701
6 0.893634678040802
};
\addplot [black, forget plot]
table {%
6 0.896953341114642
6 0.897859861721293
};
\addplot [black, forget plot]
table {%
5.875 0.893634678040802
6.125 0.893634678040802
};
\addplot [black, forget plot]
table {%
5.875 0.897859861721293
6.125 0.897859861721293
};
\addplot [semithick, color2]
table {%
0.5 0.908827448042987
6.5 0.908827448042987
};
\addlegendentry{Clean}
\addplot [semithick, color3]
table {%
0.5 0.836633657524875
6.5 0.836633657524875
};
\addlegendentry{Interfered}
\path [draw=black]
(axis cs:0.75,0)
--(axis cs:1.25,0)
--(axis cs:1.25,0)
--(axis cs:0.75,0)
--(axis cs:0.75,0)
--cycle;
\path [draw=black]
(axis cs:1.75,0.724102604931597)
--(axis cs:2.25,0.724102604931597)
--(axis cs:2.25,0.727548151004027)
--(axis cs:1.75,0.727548151004027)
--(axis cs:1.75,0.724102604931597)
--cycle;
\path [draw=black]
(axis cs:2.75,0.860000051320365)
--(axis cs:3.25,0.860000051320365)
--(axis cs:3.25,0.864417985222416)
--(axis cs:2.75,0.864417985222416)
--(axis cs:2.75,0.860000051320365)
--cycle;
\path [draw=black]
(axis cs:3.75,0.882029383878705)
--(axis cs:4.25,0.882029383878705)
--(axis cs:4.25,0.88452925871811)
--(axis cs:3.75,0.88452925871811)
--(axis cs:3.75,0.882029383878705)
--cycle;
\path [draw=black]
(axis cs:4.75,0.892895973099222)
--(axis cs:5.25,0.892895973099222)
--(axis cs:5.25,0.894190336083145)
--(axis cs:4.75,0.894190336083145)
--(axis cs:4.75,0.892895973099222)
--cycle;
\path [draw=black]
(axis cs:5.75,0.894994219860701)
--(axis cs:6.25,0.894994219860701)
--(axis cs:6.25,0.896953341114642)
--(axis cs:5.75,0.896953341114642)
--(axis cs:5.75,0.894994219860701)
--cycle;
\addplot [color1, forget plot]
table {%
0.75 0
1.25 0
};
\addplot [semithick, color1, forget plot]
table {%
1.75 0.724137727704655
2.25 0.724137727704655
};
\addplot [semithick, color1, forget plot]
table {%
2.75 0.863658382000304
3.25 0.863658382000304
};
\addplot [semithick, color1, forget plot]
table {%
3.75 0.882394420286227
4.25 0.882394420286227
};
\addplot [semithick, color1, forget plot]
table {%
4.75 0.893130328375094
5.25 0.893130328375094
};
\addplot [semithick, color1, forget plot]
table {%
5.75 0.89669028731551
6.25 0.89669028731551
};
\end{axis}

\end{tikzpicture}
		\vspace{-5mm}
		\caption{F1-Score and memory requirements for models with multiple bits per weight and activation. The model has $L=3$ layers and $C=[16, 8, 2]$ channels. Details to memory requirements, performance and bit-widths are listed in Table~\ref{tab:1d}.}
		\label{fig:result1e}
	}
\end{figure*}

\begin{table*}
	\centering
	\scriptsize
	\caption{Memory, performance and bit-width details for results shown in Figure~\ref{fig:result1e}. The memory is stated in kilobytes and the F1-Score is listed as 'mean $\pm$ standard deviation' over three independently trained models.}
	\label{tab:1d}
	\ra{1.05}
		\begin{tabular}{@{}ccrrcrrcrcr@{}}\toprule
			\multicolumn{1}{c}{Quantization} & \phantom{abc} & \multicolumn{2}{c}{Weights} & \phantom{abc}& \multicolumn{2}{c}{Feature-Maps} &
			\phantom{abc} & \multicolumn{1}{c}{Total} & \phantom{abc} & \multicolumn{1}{c}{F1}\\
			\cmidrule{3-4} \cmidrule{6-7}
			&& [kB] & Bits && [kB] & Bits && [kB] && \\ \midrule
			4 Bits && $0.773$ & 4 && $108.00$ & 4 && $108.94$ && $0.8617 \pm 0.004$ \\
			6 Bits && $1.160$ & 6 && $162.00$ & 6 && $163.32$ && $0.8836 \pm 0.002$ \\
			8 Bits && $1.547$ & 8 && $216.00$ & 8 && $217.71$ && $0.8937 \pm 0.001$ \\
			32 Bits (real-valued) && $6.188$ & 32 && $864.00$ & 32 && $870.35$ && $0.8960 \pm 0.002$ \\
			\bottomrule
		\end{tabular}
\end{table*}

In the next experiment, we aim to further reduce the memory size of these small real-valued models without a significant performance degradation. We choose the model with three layers and $C=[16,8,2]$ channels as a base model. Figure~\ref{fig:result1e} shows the quantization performance with different bit-widths (1, 2, 4, 6, 8, 32) for weights and activations alike. We use integer quantization and calculate the dynamic range as the maximum absolute value over the real-valued auxiliary weights. See Table~\ref{tab:1d} for details of memory reduction and F1-Scores.

Models with a fixed number of 1 or 2 bits are not suited for the task and do not even reach the F1-Score of signals without mitigation. With 4, 6, and 8 bits the performance increases steadily and almost reaches the real-valued score with only 8 bits. The resulting memory saving with 8-bit weights and activations is approximately 75 \% compared to the real-valued baseline.

\section{Conclusion}
In this paper, we investigate the capability to quantize CNN-based models for denoising and interference mitigation of radar signals. Our experiments show, that the initial model size and architecture have a substantial contribution to the quantization outcome, thus we emphasize the importance of small initial real-valued models in order to obtain memory efficient models after quantization.

We find that small architectures are not suitable for binarization in the context of the considered regression task and instead multiple bits are required to retain high performance. For the considered task and selected base models, the quantization of activations has a substantially higher impact on the overall memory than the quantization of weights. An 8-bit model can be used for the considered task reaching a memory reduction of approximately 75 \% compared to the real-valued equivalent without noteworthy performance degradation.

In the future, we want to analyze quantization using trained bit-width in detail and explore potential advantages of different quantization techniques.

\section*{Acknowledgments}
This work was supported by the Austrian Research Promotion Agency (FFG) under the project SAHaRA (17774193) and NVIDIA by providing GPUs.

%
%
%
 \bibliographystyle{ieeetr}
 \bibliography{mybibliography}

\end{document}